\documentclass[%
 reprint,
superscriptaddress,
 amsmath,amssymb,
 aps,
 prl,
]{revtex4-1}

\makeatletter
\newsavebox{\@brx}
\newcommand{\llangle}[1][]{\savebox{\@brx}{\(\m@th{#1\langle}\)}%
  \mathopen{\copy\@brx\kern-0.5\wd\@brx\usebox{\@brx}}}
\newcommand{\rrangle}[1][]{\savebox{\@brx}{\(\m@th{#1\rangle}\)}%
  \mathclose{\copy\@brx\kern-0.5\wd\@brx\usebox{\@brx}}}
\makeatother

\usepackage{graphicx}
\usepackage[normalem]{ulem}
\usepackage{epstopdf}
\usepackage{xcolor}
\usepackage{dcolumn}
\usepackage{bm}
\usepackage{hyperref}

\begin{document}

\title{Geometric Origin of Intrinsic Spin Hall Effect in an Inhomogeneous Electric Field}

\author{Anwei Zhang}
\email{zawcuhk@gmail.com}
\affiliation{Department of Physics, Ajou University, Suwon 16499, Korea}
\author{Jun-Won Rhim}
\email{jwrhim@ajou.ac.kr}
\affiliation{Department of Physics, Ajou University, Suwon 16499, Korea}
\affiliation{Research Center for Novel Epitaxial Quantum Architectures, Department of Physics, Seoul National University, Seoul, 08826, Korea}

\begin{abstract}
In recent years, the spin Hall effect has received great attention because of its potential application in spintronics and quantum information processing and storage. However, this effect is usually studied under the external homogeneous electric field. Understanding how the inhomogeneous electric field affects the spin Hall effect is still lacking. Here, we investigate a two-dimensional two-band time-reversal symmetric system and give an expression for the intrinsic spin Hall conductivity in the presence of the inhomogeneous electric field, which is shown to be expressed through gauge-invariant geometric quantities. On the other hand, when people get physical intuition on transport phenomena from the wave packet, one issue appears. It is shown that the conductivity obtained from the conventional wave packet approach cannot be fully consistent with the one predicted by the Kubo-Greenwood formula. Here, we attempt to solve this problem.
\end{abstract}

\maketitle

\noindent{\bfseries Introduction} The spin Hall effect (SHE) is a spin-accumulation phenomenon on the boundaries of a 2D system caused by the spin-dependent transverse deflection of the charge current~\cite{hirsch1999spin,zhang2000spin,kato2004observation,wunderlich2005experimental}.
This phenomenon has received significant attention because it can be applied to spintronics by offering a core mechanism for the generation and detection of spin current~\cite{niimi2015reciprocal,sinova2015spin}.
Depending on the origin of the SHE, it is categorized into the intrinsic and extrinsic SHE.
While the relativistic spin-orbit coupling (SOC) plays a crucial role in both cases, the intrinsic SHE  arises from the intrinsic band structure, whereas the extrinsic one is due to the impurities with large SOC~\cite{murakami2003dissipationless,murakami20042,sinova2004universal,vzutic2004spintronics}.
The intrinsic SHE has been of great interest because its underlying mechanism is irrelevant to the random impurities unlike the extrinsic case and the giant spin Hall conductivity (SHC) of several materials such as Pt is presumed to be originating from this effect~\cite{sinova2015spin,bercioux2015quantum,baltz2018antiferromagnetic,sinitsyn2004spin,shen2004resonant,erlingsson2005spin,shekhter2005chiral,yao2005sign,guo2008intrinsic,tanaka2008intrinsic,kontani2009giant,morota2011indication,werake2011observation,patri2018theory,zhu2019variation,shin2019unraveling,jadaun2020rational}.

While the previous works are mostly performed in spatially uniform fields, it has been noted that the application of nonuniform fields could lead to a variety of new phenomena~\cite{zhang2019tunable,zhang2018frozen}, and even offer us access to various geometric quantities of Bloch wave functions.
In an inhomogeneous electric field, the Hall conductivity is related to the Hall viscosity for Galilean invariant systems~\cite{bradlyn2012kubo,hoyos2012hall,holder2019unified}, the semiclassical equations of motion gain corrections depending on quantum metric~\cite{lapa2019semiclassical}, and the intrinsic anomalous Hall conductivity (AHC) is expressed through quantum metric, Berry curvature, and fully symmetric rank-$3$ tensor~\cite{kozii2021intrinsic}.
Moreover, the nonreciprocal directional dichroism, i.e., the difference of the refractive index between counterpropagating light fields, is connected with quantum metric dipole~\cite{gao2019nonreciprocal}.

In this paper, we investigate the intrinsic SHC in the inhomogeneous field. 
We consider a two-dimensional two-band system respecting time-reversal symmetry to suppress the anomalous Hall effect.
Using the Kubo-Greenwood formula, we obtain the leading correction to the conventional intrinsic SHC in the uniform electric field. 
We show that such a leading term, which is the square term of the electric field wave vector, depends on band velocity as well as gauge-invariant geometric quantities: quantum metric and interband Berry connection. 
For Rashba and Dresselhaus systems, we show that the SHC under a nonuniform field is no longer a universal value. 
Instead, it can be adjusted by tuning the Fermi energy of the system which allows us to manipulate the spin current.

In this work, we also address an issue about the incompatibility between the Kubo-Greenwood formula and the semiclassical wave packet approach revealed in the anomalous Hall effect~\cite{kozii2021intrinsic}.
To this end, we expand the perturbed Hamiltonian in terms of vector potential rather than the scalar potential and construct the wave packet by the superposition of wave functions in the upper and lower bands rather than the single lower band.
We show that thus obtained wave packet yields the SHC and AHC consistent with the Kubo-Greenwood formula.


\smallskip

\begin{figure*}[t]
\centering
\includegraphics[width=0.4\textwidth ]{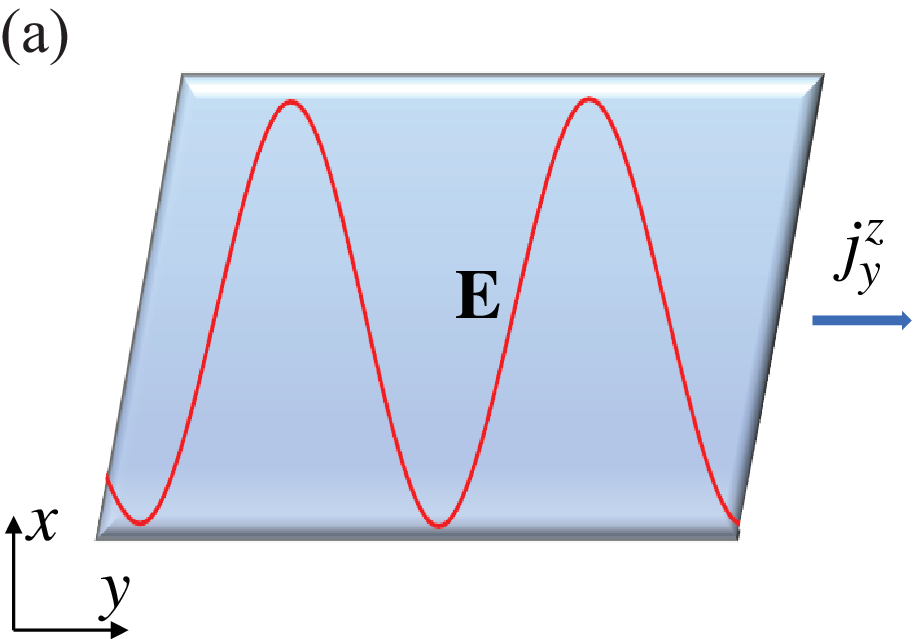}~~~~~~~~~~~
\includegraphics[width=0.4\textwidth ]{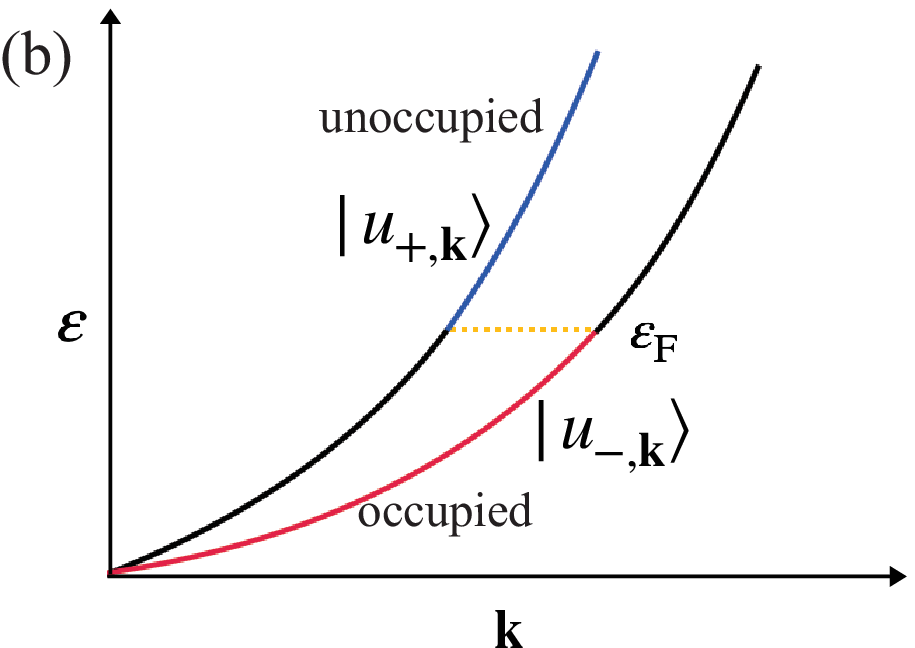}
\caption{{\bfseries Schematic of the two-dimensional two-band system.} (a) A two-dimensional system under consideration, where the inhomogeneous in-plane electric field $\mathbf{E}$ is applied. Spin Hall current $j^z_y$ flows perpendicular to the polarization direction of the applied field. (b) Energy bands model with the Fermi energy $\mathrm{\varepsilon_F}$ higher than the nodal energy. The temperate is taken as zero. $\vert u_{\pm, \mathbf{k}}\rangle$ are the periodic part of Bloch wave functions for the upper unoccupied band and the lower occupied band, respectively.
}\label{figure.1}
\end{figure*}

\noindent{\bfseries Results and Discussion}\\
\noindent{\bfseries Two-band model} We consider the two-dimensional two-band Hamiltonian 
\begin{equation}\label{eq:hamiltonian_0}
H_0=\frac{\hbar^2 k^2}{2m}+\sum^2_{i=1} d_i(\mathbf{k}) \sigma_i,
\end{equation}
where $k=\vert \mathbf{k}\vert=\sqrt{k^2_x+k^2_y}$ is the modulus of the electron momentum, $\sigma_i$ is the Pauli matrix, and $d_i(\mathbf{k})$ is an arbitrary real function. 
The two eigenenergies of this Hamiltonian are given by $\varepsilon_{\pm,\mathbf{k}}=\hbar^2 k^2/2m\pm d(\mathbf{k})$, where $d(\mathbf{k})=\sqrt{d^2_1(\mathbf{k})+d^2_2(\mathbf{k})}$.
The corresponding Bloch wave functions are obtained as $ e^{i\mathbf{k}\cdot\mathbf{r}}\left(\pm(d_1(\mathbf{k})-i d_2(\mathbf{k}))/d, 1\right)^{ \mathrm{T}}/\sqrt{2}$. 
We impose the restriction $d_i(\mathbf{k}) =-d_i(-\mathbf{k})$ to reflect time-reversal symmetry of our system. 
The Rashba and Dresselhaus Hamiltonians, representative time-reversal symmetric models, satisfy this relationship.
Under such a restriction, the eigenvalues of the system are even functions of the momentum $\mathbf{k}$.

\noindent{\bfseries Spin Hall conductivity} In this paper, we consider in-plane electric field polarized along the $x$-direction and propagating along the $y$-direction as illustrated in Fig.~\ref{figure.1}(a), i.e., $\mathbf{E}=E_x\hat{x} e^{i q y-i \omega t}$+c.c. \cite{marder2000condensed}.
Then the induced transverse spin Hall conductivity in the static limit $\omega \rightarrow 0$ is given by the Kubo-Greenwood formula
\begin{equation}
\sigma_{\mathrm{SH}}(\mathbf{q}) = \frac{-2e\hbar}{S}\sum_{\mathbf{k}}\frac{\mathrm{Im}[\langle \hat{j}^z_y \rangle_{\mathbf{k},\mathbf{q}}\langle \hat{v}_x \rangle^*_{\mathbf{k},\mathbf{q}}]}{\Delta\varepsilon^2_{\mathbf{k},\mathbf{q}}},\label{eq:cond}
\end{equation}
where $-e$ is the charge of the electron, $S$ is the area of the system.
We define $\langle \hat{O} \rangle_{\mathbf{k},\mathbf{q}} \equiv \langle u_{-,\mathbf{k}-\mathbf{q}/2}\vert \hat{O}\vert u_{+,\mathbf{k}+\mathbf{q}/2}\rangle$ and $\Delta\varepsilon_{\mathbf{k},\mathbf{q}} \equiv \varepsilon_{-, \mathbf{k}-\mathbf{q}/2}-\varepsilon_{+, \mathbf{k}+\mathbf{q}/2}$, where $\vert u_{\pm, \mathbf{k}}\rangle$ are the periodic part of Bloch wave functions for the upper unoccupied band $(+)$ and the lower occupied band $(-)$ respectively, as shown in Fig.~\ref{figure.1}(b).
Here $\mathbf{q}=q\hat{y}$ represents the momentum transfer due to the modulation of the electric field, $\hat{v}_x=\partial_{k_x}H_0/\hbar$ is the velocity operator along $x$-direction, and $\hat{j}^z_y=\hbar\{ \sigma_3, \hat{v}_y\}/4=(\hbar^2 k_y/2m)\sigma_3$ is the spin current operator polarized perpendicular to the $xy$-plane and flowing alone the $y$-direction.

In the uniform field limit ($q \rightarrow 0$), we have the conventional intrinsic SHC
\begin{equation}\label{eq:sigma_0}
\sigma^{(0)}_{\mathrm{SH}}=\frac{-e\hbar^2}{m S}\sum_{\mathbf{k}}\frac{k_y}{\Delta\varepsilon_{\mathbf{k}}}A^x_{+-},
\end{equation}
where $A^x_{+-}=i\langle u_{+,\mathbf{k}} \vert \partial_{k_x}\vert u_{-,\mathbf{k}}\rangle$ is the interband or cross-gap Berry connection~\cite{de2017quantized,hwang2021geometric} along $x$-axis and $\Delta\varepsilon_{\mathbf{k}} \equiv \Delta\varepsilon_{\mathbf{k},0} = \varepsilon_{-,\mathbf{k}}-\varepsilon_{+,\mathbf{k}}$ denotes the energy difference between two bands at a given momentum. 
Here the relation $\langle u_{-,\mathbf{k}} \vert \sigma_3\vert u_{+,\mathbf{k}}\rangle=-1$ is used.

 \begin{figure*}[t]
\centering
\includegraphics[width=0.4\textwidth ]{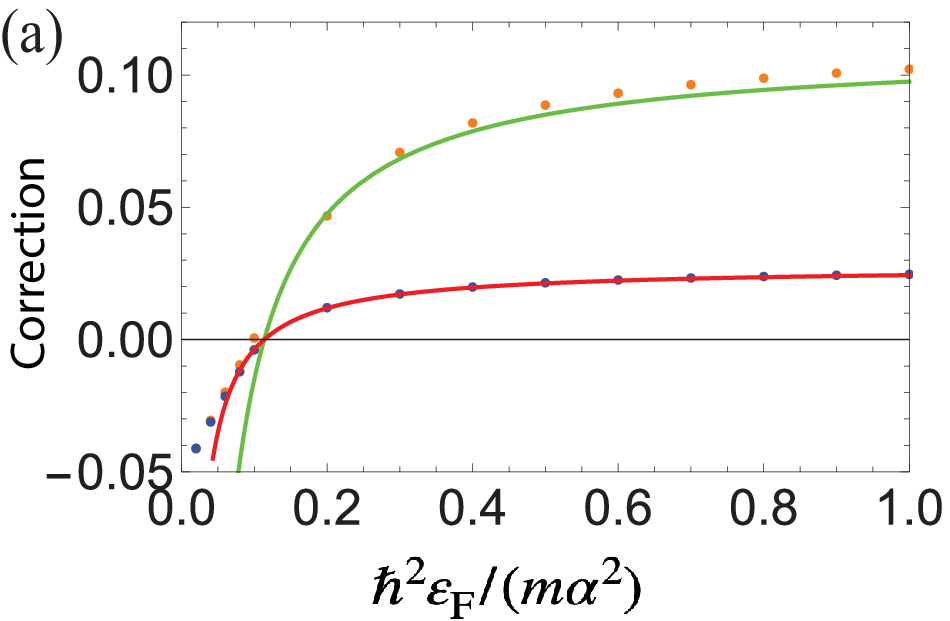}~~~~~~~~~~~~
\includegraphics[width=0.4\textwidth ]{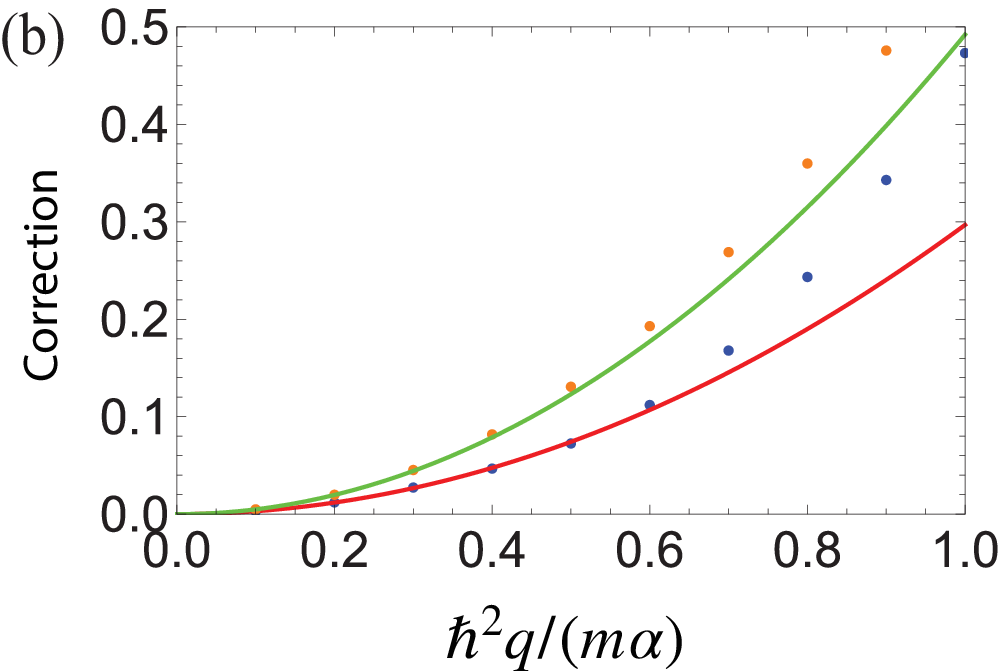}
\caption{{\bfseries The correction in unit of $e/(8\pi)$ to the spin Hall conductivity of the Rashba system as a function of (a) the dimensionless Fermi energy $\hbar^2\mathrm{\varepsilon_F}/(m\alpha^2)$ and (b) the dimensionless wave vector $\hbar^2q/(m\alpha)$, respectively.} 
In (a), the green and red solid lines are the correction $q^2\sigma^{(2)}_{y x}$ with $\hbar^2q/(m\alpha)=0.4$ and $\hbar^2q/(m\alpha)=0.2$, respectively.
In (b), the green and red solid lines are the correction $q^2\sigma^{(2)}_{y x}$ with $\hbar^2\mathrm{\varepsilon_F}/(m\alpha^2)=0.4$ and $\hbar^2\mathrm{\varepsilon_F}/(m\alpha^2)=0.2$, respectively.
The dotted lines refer to the total correction with the corresponding parameters.
}\label{figure.2}
\end{figure*}

For nonuniform external electric field with long wavelength
excitation, the SHC can be expanded as $\sigma_{\mathrm{SH}}(q) =\sigma^{(0)}_{\mathrm{SH}} + q\sigma^{(1)}_{\mathrm{SH}}+q^2\sigma^{(2)}_{\mathrm{SH}}+ O (q^3)$.
Due to the time-reversal symmetry of our system, the band velocity and interband Berry connection are odd functions of $\mathbf{k}$.
As a result, the $q$-linear term vanishes after integration over $\mathbf{k}$.
Then the $q^2$ term becomes the leading term for the deviation of the SHC from its value under uniform electric field, which is given by (see Supplementary Methods)
\begin{widetext}
\begin{eqnarray} \label{eq:sigma_2}
\sigma^{(2)}_{\mathrm{SH}}&=&\frac{e\hbar^2}{4m S}\sum_{\mathbf{k}}k_y\bigg[\frac{2g_{yy}}{\Delta\varepsilon_{\mathbf{k}}}A^x_{+-}+\frac{\partial_{k_y}(v_{-,y}-v_{+,y})}{\Delta\varepsilon^2_{\mathbf{k}}}A^x_{+-}+\frac{v_{-,x}-v_{+,x}}{2\Delta\varepsilon^2_{\mathbf{k}}}\partial_{k_y} A^y_{+-}\nonumber\\&&+\frac{2\hbar^2(v_{-,x}+v_{+,x})(v_{-,y}+v_{+,y})}{\Delta\varepsilon^3_{\mathbf{k}}} A^y_{+-}-\frac{3\hbar^2(v_{-,y}+v_{+,y})^2 }{\Delta\varepsilon^3_{\mathbf{k}}} A^x_{+-}\bigg],
\end{eqnarray}
\end{widetext}
where $v_{\pm,i}=\partial_{k_i}\varepsilon_{\pm,\mathbf{k}}/\hbar$ is the group velocity along $i$-axis ($i=x,y$), $A^y_{+-}=i\langle u_{+,\mathbf{k}} \vert \partial_{k_y}\vert u_{-,\mathbf{k}}\rangle$ is the interband Berry connection with respective to $k_y$, and $g_{yy}$ is the Fubini-Study quantum metric \cite{provost1980riemannian} for the filled band given by
\begin{equation}\label{5}
g_{yy}= \mathrm{Re}[\langle \partial_{k_y} u_{-,\mathbf{k}} \vert u_{+,\mathbf{k}} \rangle \langle u_{+,\mathbf{k}} \vert \partial_{k_y} u_{-,\mathbf{k}}\rangle].
\end{equation}
Here we have used the relation $\langle u_{\pm,\mathbf{k}} \vert \sigma_3\vert u_{\pm,\mathbf{k}}\rangle=0$, which means that spin-up and -down are equally distributed.
 
Under the gauge transformation $\vert u_{\pm,\mathbf{k}}\rangle \rightarrow e^{i\phi_\mathbf{k}}\vert u_{\pm,\mathbf{k}}\rangle$, both the interband Berry connection and quantum metric are invariant.
Therefore, the correction in (\ref{eq:sigma_2}) is also gauge-invariant. 
If we impose a further restriction for our system: $d_{j}(\mathbf{k})$ is a linear function of the momentum, i.e., $\partial^2_{k_i}d_{j}(\mathbf{k})=0$, our result (\ref{eq:sigma_2}) can be further simplified because $\partial_{k_i}(v_{-,i}-v_{+,i})=4\Delta\varepsilon_{\mathbf{k}} g_{ii}$ in such a case~\cite{zhang2021revealing}.

The theory can be extended to more general case. When the term $\hbar^2 k^2/2m$ in the Hamiltonian (\ref{eq:hamiltonian_0}) is replaced by $d_0(\mathbf{k})$ which satisfies
the restriction $d_0(\mathbf{k})=d_0(-\mathbf{k})$, the spin current operator will become $\hat{j}^z_y=\partial_{k_y}{d_0(\mathbf{k})}\sigma_3/2$. As a  consequence, the factor $\hbar^2 k_y/m $ in the results (\ref{eq:sigma_0}) and (\ref{eq:sigma_2}) is simply changed to $\partial_{k_y}{d_0(\mathbf{k})}$.

\noindent{\bfseries Rashba and Dresselhaus models} Let us consider the Rashba model given by
$H_{0}^\mathrm{R}=\hbar^2 k^2/2m+\alpha k_y  \sigma_x-\alpha k_x  \sigma_y$, where $\alpha$ is the strength of the Rashba SOC. 
For this system, we have interband Berry connection $A^x_{+-}=k_y/2k^2$, $A^y_{+-}=-k_x/2k^2$, and quantum metric $g_{yy}=k^2_x/4k^4$.
Using (\ref{eq:sigma_0}) and (\ref{eq:sigma_2}), the SHC up to $q^2$-term is evaluated as
\begin{equation}\label{eq:sigma_rashba}
\sigma^{\mathrm{R}}_{\mathrm{SH}}(q)=\frac{e}{8\pi}\bigg[1+\frac{\hbar^4q^2}{16m^2\alpha^2}\bigg(11-\frac{5m\alpha^2}{4\hbar^2\mathrm{\varepsilon_F}}\bigg)\bigg],
\end{equation}
where $\mathrm{\varepsilon_F}$ is the Fermi energy.
As shown in Fig.~\ref{figure.2}(a), one can adjust the intrinsic SHC by tuning the Fermi energy, unlike the uniform electric field case. This provides us with a new way to manipulate spin current. It is worth noting that in the homogeneous field case, SHC in other systems can be tunable ~\cite{moca2007spin,csahin2015tunable,li2021tunable}.
The spin Hall conductivity (\ref{eq:sigma_rashba}) obtained from the geometric formula (\ref{eq:sigma_2}) matches well with the exact result from (\ref{eq:cond}) as $\mathrm{\varepsilon_F}$ grows and $q$ decreases as plotted in Fig.~\ref{figure.2}(a) and (b). 
This is due to the fact that the perturbation method for expanding the conductivity is valid for small $q/\Delta\varepsilon^2_{\mathbf{k}}$.

If we perform the transformation: $d_1(\mathbf{k})\leftrightarrow d_2(\mathbf{k})$ and $\alpha\rightarrow \beta$, the Rashba system changes to the Dresselhaus model with the Hamiltonian
$H_0^\mathrm{D}=\hbar^2 k^2/2m-\beta k_x  \sigma_x+\beta k_y  \sigma_y$, where $\beta$
is the Dresselhaus coupling strength. 
Under such transformation, the band structure and quantum metric remain the same, while the interband Berry connection changes its sign: $A^{j}_{+-}\rightarrow -A^{j}_{+-}$.
As a result, the SHC of the Dresselhaus system becomes
\begin{equation}\label{eq:sigma_dresselhaus}
\sigma_{\mathrm{SH}}^{\mathrm{D}}(q)=-\frac{e}{8\pi}\bigg[1+\frac{\hbar^4q^2}{16m^2\beta^2}\bigg(11-\frac{5m\beta^2}{4\hbar^2 \mathrm{\varepsilon_F}}\bigg)\bigg].
\end{equation}
There is a sign different from the result of the Rashba model. 

There are many materials, whose electronic structures are descried by the Rashba or Dresselhaus model, such as (i) 2D interfaces of InAlAs/InGaAs~\cite{nitta1997gate} and LaAlO$_3$/SrTiO$_3$~\cite{herranz2015engineering}, (ii) Si metal-oxide- semiconductor (MOS) heterostructure~\cite{lee2021synthetic}, and (iii) surfaces of heavy materials like Au~\cite{lashell1996spin} and BiAg(111)~\cite{ast2007giant,hong2019giant}.
By using the well-known experimental techniques probing intrinsic spin Hall effect~\cite{murakami2003dissipationless,sinova2004universal,wunderlich2005experimental,valenzuela2006direct,choi2015electrical}, we expect to detect the intriguing Fermi level-dependence of the $q^2$-term of the SHC under the spatially modulating electric field by tuning its wavelength.

Apart from Rashba and Dresselhaus system, the theory is also applicable to the two-dimensional heavy holes in III-V semiconductor quantum wells with the cubic Rashba coupling~\cite{winkler2000rashba,schliemann2005spin}: 
$H_0^\mathrm{C}=\hbar^2 k^2/2m+i\alpha(k^3_-\sigma_+-k^3_+\sigma_-)/2$, where $k_{\pm}=k_x \pm ik_y$ and $\sigma_{\pm}=\sigma_x \pm i\sigma_y$. In such a material system, $d_1(\mathbf{k})$ and $d_2(\mathbf{k})$ are respectively $\alpha k_y(3k^2_x-k^2_y)$ and $\alpha k_x(3k^2_y-k^2_x)$, which can reflect time-reversal symmetry. Here we should note that the angular momentum quantum numbers of the heavy holes is $3/2$, so the spin current operator is $\hat{j}^z_y=3(\hbar^2 k_y/2m)\sigma_3$~\cite{schliemann2005spin} and the results (\ref{eq:sigma_0}) and (\ref{eq:sigma_2}) should be multiplied by $3$.

\noindent{\bfseries Wave packet approach}  Conventionally, people get physical intuition for transport phenomena from the semiclassical analysis conducted based on the wave packet dynamics.
However, it was mentioned that the AHC calculated from the single-band wave packet method could be inconsistent with the one predicted by the Kubo-Greenwood formula~\cite{kozii2021intrinsic}. 
We show that we should construct the wave packet from two bands for the AHC or SHC to be consistent with the Kubo-Greenwood formula.

Under the presence of the external field, the Hamiltonian of the system is  $H=H_0+H^{'}$. Here $H^{'}$ is the perturbative coupling term with the field, which is given by
\begin{equation}\label{8}
H^{'}=e\hat{v}_x A_x,
\end{equation}
where $A_x=(E_x/i\omega) e^{i q y-i \omega t}$+c.c. is the vector potential. It's worth noting that in order to get the transverse conductivity, we expand the perturbed Hamiltonian in terms of the vector potential rather than scalar potential \cite{dressel_grner_2002}.

A wave packet is usually constructed from the unperturbed Bloch wave function within a single band.
However, this conventional way gives inconsistent results with the Kubo-Greenwood formula when calculating AHC~\cite{kozii2021intrinsic}. 
Besides, as can be seen from (\ref{eq:sigma_0}) and (\ref{eq:sigma_2}), the SHC results from the interband coupling. 
The single-band wave packet method yields a null result for the SHC. 
Therefore, we use both the upper and lower bands to construct the wave packet as
\begin{equation}\label{9}
\vert \psi_- (t) \rangle=\int d\mathbf{k} a(\mathbf{k},t)e^{i \mathbf{k} \cdot\mathbf{r}}\bigg(\vert u_{-,\mathbf{k}-\frac{\mathbf{q}}{2}}\rangle+\frac{H^{'}_{+-}}{\Delta\varepsilon_{\mathbf{k},\mathbf{q}}}\vert u_{+,\mathbf{k}+\frac{\mathbf{q}}{2}}\rangle\bigg),
\end{equation}
where $H^{'}_{+-}=\langle u_{+,\mathbf{k}+\mathbf{q}/2}\vert H^{'}\vert u_{-,\mathbf{k}-\mathbf{q}/2}\rangle$ is the transition matrix element, and the amplitude $a(\mathbf{k},t)$ satisfies the normalization condition $\int d\mathbf{k} \vert a(\mathbf{k},t)\vert^2=1$. 
One can note that the Bloch wave functions in the upper band are involved in constructing the wave packet in the same manner as the first-order stationary perturbation scheme for the lower band.
A similar wave packet structure can be found in Ref.~\cite{gao2014field,gao2019nonreciprocal} and in the non-Abelian formulation~\cite{xiao2010berry}.

The spin current flowing  alone $y$-direction can be evaluated as
\begin{equation}\label{spin current}
j^z_y= \frac{\hbar}{2S}\sum_{\mathbf{k_c}}(\dot{R}_{y\uparrow}-\dot{R}_{y\downarrow}),
\end{equation}
where $\mathbf{k_c}$ is the momentum of the wave packet,  and $R_{y\sigma}=\langle \psi_{-\sigma} (t) \vert \hat{r}_y\vert \psi_{-\sigma}(t) \rangle$ is the average position of the spin-$\sigma$ part of the wave packet given by $\vert \psi_{-\uparrow}(t) \rangle=(1,0)\vert \psi_- (t) \rangle$ and $\vert \psi_{-\downarrow}(t) \rangle=(0,1)\vert \psi_- (t) \rangle$ for spin-up and -down, respectively. 
Since the expression $\langle \psi_{-\sigma} (t) \vert \hat{O}\vert \psi_{-\sigma}(t) \rangle$ is actually $\langle \psi_- (t) \vert 2\hat{O}+\mathrm{sgn}(\sigma)\{\sigma_3, \hat{O}\}\vert \psi_- (t) \rangle/4$, we have
\begin{eqnarray}\label{velo}
  \dot{R}_{y\sigma}&=&\langle \psi_- (t) \vert \hat{v}_y/2+\mathrm{sgn}(\sigma)\hat{j}^z_y/\hbar\vert \psi_- (t) \rangle,
\end{eqnarray}
where $\mathrm{sgn}(\sigma)=1(-1)$ for spin-up(down).
It is worth to note that in this paper, we use the classical form of the spin current operator $\hat{j}^z_y=\hbar\{ \sigma_3, \hat{v}_y\}/4$~\cite{sinova2004universal,shin2019unraveling} to substitute the effective spin current operator $(\hbar/2) d(\hat{r}_y\sigma_3)/dt$~\cite{shi2006proper}. Besides,
for time-reversal symmetric system, there is no anomalous Hall effect, i.e., $\langle \psi_- (t) \vert \hat{v}_y\vert \psi_- (t) \rangle=0$. Therefore in such a case, the velocities in spin-up and -down basis satisfy the relation: $ \dot{R}_{y\uparrow}=-\dot{R}_{y\downarrow} =\langle \psi_- (t) \vert \hat{j}^z_y\vert \psi_- (t) \rangle/\hbar$, which indicates that the spin-up and -down part of wave packet have opposite velocity.
On the other hand, for time-reversal symmetry breaking system, the charge current along $y$-direction is 
\begin{equation}\label{char}
j_y= \frac{-e}{S}\sum_{\mathbf{k_c}}\dot{R}_y=\frac{-e}{S}\sum_{\mathbf{k_c}}(\dot{R}_{y\uparrow}+\dot{R}_{y\downarrow}),
\end{equation}
where $R_y=\langle \psi_- (t) \vert \hat{r}_y\vert \psi_- (t) \rangle=R_{y\uparrow}+R_{y\downarrow}$ is the position of the wave packet. 

The spin current and charge current can be written in the same form $j= \frac{1}{S}\sum_{\mathbf{k_c}}\langle \psi_- (t) \vert \hat{j}\vert \psi_- (t) \rangle$, where $\hat{j}$ is $\hat{j}^z_y$ for spin current or $\hat{j}_y=-e\hat{v}_y$ for charge current. 
Substituting (\ref{9}) into it, we have
\begin{equation}\label{11}
j=\frac{1}{S}\sum_{\mathbf{k_c}}\int d\mathbf{k}a^2(\mathbf{k},t)\frac{\langle u_{-,\mathbf{k}-\frac{\mathbf{q}}{2}}\vert \hat{j}\vert u_{+,\mathbf{k}+\frac{\mathbf{q}}{2}}\rangle H^{'}_{+-}}{\Delta\varepsilon_{\mathbf{k},\mathbf{q}}}+\mathrm{c.c.}.
\end{equation}
According to Fermi's golden rule, we take $\omega$ as $-\Delta\varepsilon_{\mathbf{k},\mathbf{q}}/\hbar$ for the transition matrix element  $H^{'}_{+-}$ and $\Delta\varepsilon_{\mathbf{k},\mathbf{q}}/\hbar$ for the transition matrix element $H^{'}_{-+}$. Then from
(\ref{8}) and (\ref{11}), the conductivity $j/(E_xe^{i q y-i \omega t})$ can be obtained as
\begin{equation}\label{eq:current}
\sigma(\mathbf{q}) = \frac{-2e\hbar}{S}\sum_{\mathbf{k_c}}\frac{\mathrm{Im}[\langle \hat{j} \rangle_{\mathbf{k_c},\mathbf{q}}\langle \hat{v}_x \rangle^*_{\mathbf{k_c},\mathbf{q}}]}{\Delta\varepsilon^2_{\mathbf{k_c},\mathbf{q}}},
\end{equation}
where $\sigma(\mathbf{q})$ represents the SHC or AHC depending on the choice of $\hat{j}$.
Here we have taken $\vert a(\mathbf{k},t)\vert^2\approx \delta{(\mathbf{k}-\mathbf{k_c})}$, i.e., the wave packet is sharply peaked at the momentum $\mathbf{k_c}$.  
One can note that the Kubo-Greenwood formula is reproduced from the wave packet approach, i.e., the two-band wave packet approach is compatible with the Kubo-Greenwood formula.

\noindent{\bfseries Discussion} 
In this paper, we use vector potential to expand the perturbed Hamiltonian and then construct the wave packet. In semiclassical theory, one usually adopts scalar potential: $H^{'}_s=-e\phi(x)$. To show the link between the wave packets constructed from these two gauges, here we consider the case of $q=0$. 
Under such a case, the transition matrix element becomes $H^{'}_{+-}=\langle u_{+,\mathbf{k}}\vert H^{'}\vert u_{-,\mathbf{k}}\rangle=-ie\Delta\varepsilon_{\mathbf{k}}A^x_{+-}A_x/\hbar$, which can be written as
$H^{'}_{+-}=e E A^x_{+-}$ if we take into account of the Fermi’s golden rule. Here $E$ refers to $\mathbf{E}\cdot\hat{x}$.
As a consequence, the wave packet (\ref{9}) becomes
\begin{equation}\label{sce9}
\vert \psi_- (t) \rangle=\int d\mathbf{k} a(\mathbf{k},t)e^{i \mathbf{k} \cdot\mathbf{r}}\bigg(\vert u_{-,\mathbf{k}}\rangle+\frac{e EA^x_{+-}}{\Delta\varepsilon_{\mathbf{k}}}\vert u_{+,\mathbf{k}}\rangle\bigg),
\end{equation}
which is the same with the wave packet constructed from the perturbed Hamiltonian $H^{'}_s=-e\phi(0)+eEx+O(x^2)$ at first order approximation~\cite{gao2019nonreciprocal}.

 In this work, we mainly restrict our system to the time-reversal symmetry. 
If we relax our restrictions, the leading correction to the SHC is generally the first order term of the electric field wave vector $q$: 
\begin{equation}\label{eq:current2}
\sigma^{(1)}_{\mathrm{SH}} = \frac{-e \hbar^2 }{2m S}\sum_{\mathbf{k}}\sum_{s=\pm}k_y\left[\frac{v_{s,y}}{\Delta\varepsilon^2_{\mathbf{k}}}A^x_{+-}-\frac{v_{s,x}}{2\Delta\varepsilon^2_{\mathbf{k}}}A^y_{+-}\right], 
\end{equation}
where the interband Berry connection $A_{+-}$ is generally not an odd function of $\mathbf{k}$. 
If we break the time-reversal symmetry by adding the mass term $M \sigma_3$ to our Hamiltonian, we will have
$\sigma^{(1)}_{\mathrm{SH}} =(e \hbar^2/2m S)\sum_{\mathbf{k},s} k_y\sigma_{-+}\big[(v_{s,y}/\Delta\varepsilon^2_{\mathbf{k}})\mathrm{Re}A^x_{+-}-(v_{s,x}/2\Delta\varepsilon^2_{\mathbf{k}})\mathrm{Re}A^y_{+-}\big]$, where $\sigma_{-+}=\langle u_{-,\mathbf{k}} \vert \sigma_3\vert u_{+,\mathbf{k}}\rangle=-d(\mathbf{k})/\sqrt{d(\mathbf{k})^2+M^2}$.

Here, we investigate the system with a two-by-two Hamiltonian.  
One future direction would be to study the intrinsic SHE for the system with four-by-four Dirac Hamiltonian  $H_0=d_0(\textbf{k})+\sum^{5}_{i=1}d_i(\textbf{k}) \Gamma_i$ \cite{murakami2003dissipationless,murakami20042}, where $d_0(\textbf{k})$, $d_i(\textbf{k})$ are real functions of the momentum $\textbf{k}$, and $\Gamma_i$ are four-by-four Dirac matrices satisfying the anti-commutation relations $\{\Gamma_i, \Gamma_j\}=2\delta_{ij}$. 

\noindent{\bfseries Conclusions}\\
In summary, we have studied the intrinsic spin Hall conductivity for a two-dimensional time-reversal symmetric system under an inhomogeneous electric field. 
We derive a formula for the leading correction, which is second-order in the electric field wave vector, to the conventional intrinsic spin Hall conductivity under the uniform electric field and show that it is determined by the gauge-invariant geometric quantities: quantum metric and interband Berry connection.
We show that for Rashba and Dresselhaus systems, the inhomogeneous intrinsic spin Hall conductivity is adjustable with the Fermi energy and the electric field wave vector.
We demonstrate that the incompatibility between the conventional wave packet description and the Kubo-Greenwood formula can be addressed by the modified wave packet approach.
\\
\\
\noindent{\bfseries Acknowledgements}\\
A.Z. and J.-W.R. were supported by the National Research
Foundation of Korea (NRF) Grant funded by the Korea government
(MSIT) (Grant No. 2021R1A2C1010572). J.-W.R. was supported by
the National Research Foundation of Korea (NRF) Grant funded by
the Korea government (MSIT) (Grant No. 2021R1A5A1032996).
\\
\\
\noindent{\bfseries Data availability}\\
The data that support the findings of this study are available from the corresponding authors on reasonable request.
\\
\\
\noindent{\bfseries Competing interests}\\
The authors declare no competing interests.
\\
\\
\noindent{\bfseries Author contributions}\\
A.Z. conceived this project and did the derivation. J.-W.R. provided ideas for some parts of this work.
A.Z. and J.-W.R. wrote the manuscript.
\\


%

\end{document}